\documentclass[preprint,12pt]{elsarticle}

\usepackage{amssymb}
\usepackage{amsmath}
\usepackage{gensymb}

\usepackage{wasysym}

\usepackage{graphics}
\usepackage{graphicx}
\usepackage{color}
\usepackage{subfigure}

\usepackage{rotating}
\usepackage{textcomp}
\usepackage{subfigure}
\usepackage{xcolor}

\usepackage{amsmath,amsthm,amssymb,amsfonts}
\usepackage{units}
\usepackage{hyperref}

\usepackage{dcolumn}

\usepackage{lineno}

\usepackage{verbatim}
\usepackage{here}

\begin{document}
\begin{frontmatter}
\title{Neutron production and thermal moderation at the PSI UCN source}
\author[PSI,ETH]{H.~Becker}
\author[PSI]{G.~Bison}
\author[PSI]{B.~Blau}
\author[PSI]{Z.~Chowdhuri}
\author[PSI]{J.~Eikenberg}
\author[PSI]{M.~Fertl}
\author[PSI,ETH]{K.~Kirch}

\author[PSI]{B.~Lauss\corref{cor1}}

\author[PSI]{G.~Perret}
\author[PSI]{D.~Reggiani}
\author[PSI]{D.~Ries}
\author[PSI]{P.~Schmidt-Wellenburg}
\author[PSI]{V.~Talanov\corref{cor2}}

\author[PSI]{M.~Wohlmuther}
\author[PSI]{G.~Zsigmond}

\address[PSI]{Paul Scherrer Institute, CH-5235 Villigen PSI, Switzerland}

\address[ETH]
{Institute for Particle Physics, Eidgen\"ossische Technische Hochschule, Z\"urich,  Switzerland}

\cortext[cor1]{Corresponding author, bernhard.lauss@psi.ch}
\cortext[cor2]{Corresponding author, vadim.talanov@psi.ch}

\begin{abstract}
We report on gold foil activation measurements 
performed along a vertical channel along the tank of
the ultracold neutron source at the Paul Scherrer Institute. 
The activities obtained at various distances from the spallation target
are in very good agreement with MCNPX simulations which take into account 
the detailed description of the source as built.
\end{abstract}


\end{frontmatter}



\section{Introduction}
\label{intro}


Since 2011,
the high-intensity ultracold neutron (UCN) source \cite{Daum2009,Lauss2011,Lauss2012,Lauss2014}
is being operated
at the Paul Scherrer Institute (PSI), Villigen, Switzerland.
The UCN facility serves mainly fundamental neutron physics experiments. 
Ultracold neutrons have energies below about 300\,neV, 
corresponding to milli-Kelvin temperatures.
UCN can be stored in vessels where they can be observed for hundreds of seconds \cite{golub}.

The first experiment 
installed at the new facility is an improved search for the
neutron electric dipole moment (nEDM) \cite{PPnedmPSI}. 
The search for the nEDM is considered to be one of the most important experiments in particle physics 
today (see e.g.~\cite{Raidal2008,nedm-review,strategy2013})
and will contribute to solving the puzzle of 
the matter-antimatter asymmetry observed in our universe.
The present best nEDM limit \cite{nedm-Baker2006}
can only be improved with a higher 
experimental sensitivity 
and needs a higher 
intensity of ultracold neutrons.

The PSI UCN source uses solid deuterium to produce UCN.
The intensity of UCN provided to the experiments 
is proportional to 
the thermal neutron flux 
entering the 
solid deuterium moderator. 
The understanding of the neutron flux from the spallation target,
its thermal moderation
and its spacial and energetic variation over the UCN source volume 
is essential
for assuring the 
highest possible thermal neutron production
and maximal UCN production.
A comparable spallation-based UCN source has been operated
at the Los Alamos National Laboratories for several years
~\cite{LANL2000,LANL2013}.

In this paper we compare
results from measurements of the neutron flux
using gold foil activation and simulation of these measurements
using MCNPX \cite{Pelowitz2011}.

The outline of the paper is as follows:
First we give a short introduction to the UCN source (Sec.~\ref{UCNsource}).
We describe the neutron production simulation and the 
spallation target (Sec.~\ref{target}).
In Sec.~\ref{experiment}.
the gold foil activation measurements are discussed.
Next, the Monte Carlo simulation of the neutron flux 
using the numerical model of the UCN source is explained
(Sec.~\ref{simulation}).
The results section~\ref{results}
compares calculated activities and measurements
which are then discussed and summarized in Sec.~\ref{summary}.


\section{The ultracold neutron source}
\label{UCNsource}

\begin{figure}[htb]
\centering
\includegraphics[width=0.4\textwidth]{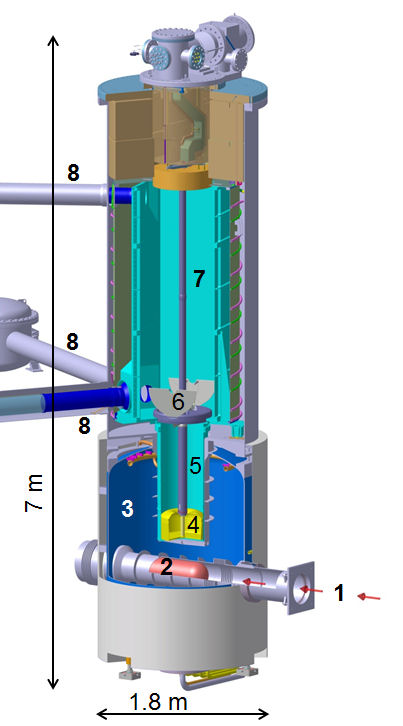}
\caption{Sketch of the UCN tank with parts important to this work: 
(1) Proton beam channel from cyclotron,
(2) spallation target,
(3) heavy water moderator tank,
(4) deuterium moderator vessel,
(5) vertical UCN guide,
(6) UCN storage vessel shutter,
(7) UCN storage vessel,
(8) UCN guides to experiments.
}
\label{fig:UCNtank}
\end{figure}

The operating principle and progress of the PSI UCN source have been documented
in \cite{Daum2009,Lauss2011,Lauss2012,Lauss2014}. A sketch of the UCN tank 
displaying the parts relevant for the investigations presented here
is given in Fig.~\ref{fig:UCNtank}.

The neutron production is based on proton induced spallation of lead. 
The 590\,MeV protons from the PSI cyclotron \cite{cyclotron2010,cyclotron2013}
are available at 2.2\,mA beam current 
24\,hours per day.
The UCN source operates
at about
1\% duty cycle with
up to 8\,s long proton beam pulses
at full beam current impinging on the target,
corresponding to
20~$\micro $A average beam current.
A typical operation with 4\,s beam pulses allows a beam pulse repetition time of 440\,s.

The resulting neutron flux off the spallation target of about 10$^{17}$ neutrons/s 
(the neutron yield is about 7~--~8 n/p \cite{Wohlmuther2006})
during the beam kick 
is thermalized by heavy water (D$_2$O)
surrounding the target at an
operating temperature of 31$^{\circ}$\,C.
Subsequently, about 30\,liters of solid deuterium inside the D$_2$ moderator vessel
at a temperature of 
about 5\,K are used 
as cold moderator and 
superthermal UCN converter \cite{Kirch2010}.
After the beam pulse the UCN are trapped in the UCN storage vessel
with its bottom shutter closed.
About 8\,m long UCN guides which penetrate the biological shielding
provide UCN to experiments
in two different experimental areas; 
the nEDM experiment is
located in area South.

\clearpage

\section{Neutron production and moderation}
\label{target}


\subsection{Spallation target}

After acquiring the final beam energy of 590\,MeV in the PSI ring cyclotron the 
full proton beam is deflected
using a kicker magnet \cite{kicker2005}
towards the UCN spallation target.
Before impinging on the target, monitors check the proton beam's center and profile. 
The profile of the proton beam at the target, as derived from the beam optics calculation, 
follows a two dimensional Gaussian distribution with 
$\sigma_{x,y}$ = 4\,cm, cut by collimators at 
a radius of
$r$=10\,cm from the beam axis \cite{Reggiani2010}. 
The corresponding intensity of the beam on target is
95$\%$ of the nominal 2.2\,mA current for the measurements regarded in this paper.

Before the system had been built 
studies of neutron yield, moderation, 
UCN production and radiation environment \cite{Atchison2002}
lead to an initial UCN source design. 
A comprehensive study of a more detailed geometry of spallation target 
and target region followed
optimizing the 
target mechanical design with respect to neutron production and 
thermo-mechanical behavior \cite{Wohlmuther2006}.

\begin{figure}[htb]
\centering
\includegraphics[width=0.4\textwidth]{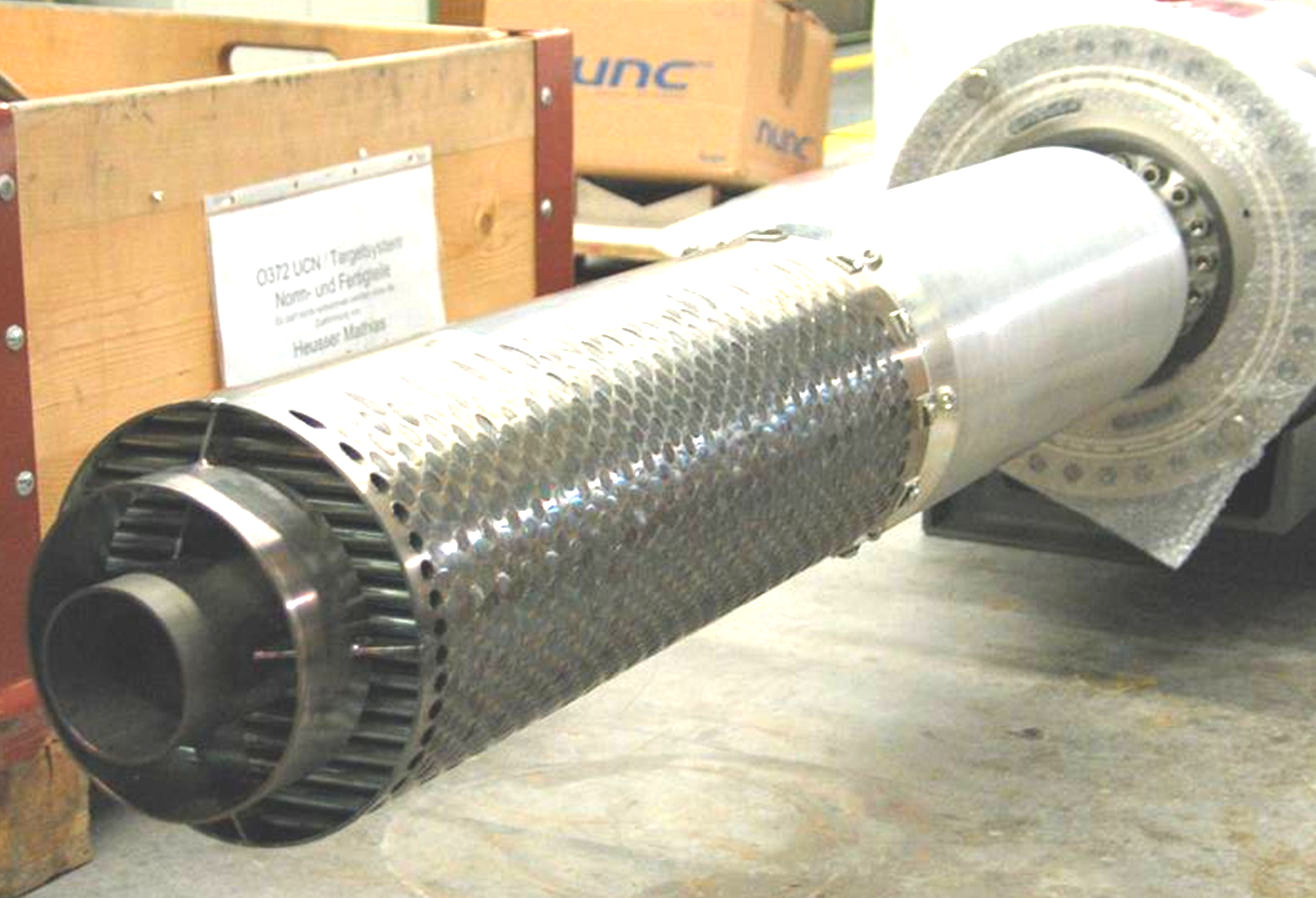}
\includegraphics[width=0.5\textwidth]{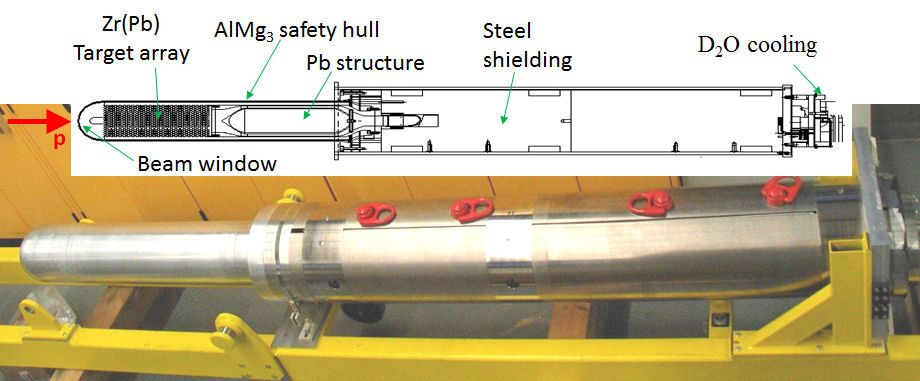}
\caption{Top: Photo of the ``Cannelloni'' array of the spallation target,
consisting of lead-filled Zircaloy tubes. 
Two cylindrical flow guides made from 2\,mm thin aluminum
direct the D$_2$O coolant flow
in front of the target.\newline
Bottom: Photo and cut view of the assembled target with safety hull ready for installation.
The total length of the target assembly is 390\,cm, the total weight about 2000\,kg.
}
\label{fig:target}
\end{figure}

The UCN source now utilizes the ``Cannelloni'' type spallation target 
assembly which has been developed and successfully 
used for a decade at PSI at the Swiss spallation neutron source SINQ \cite{Wagner2012}.
Fig.~\ref{fig:target} shows a photo of the spallation target before its installation at the UCN source. 
The target array,
with a length of 55\,cm and a radius of 10.5\,cm consists of 760 Zircaloy 
tubes with 10.75\,mm outside 
and 9.25\,mm inside 
diameter, each filled with lead at a filling factor of 90$\%$. 
The total mass in the target is 93\,kg of lead,
21\,kg of Zircaloy and 7.5\,kg of D$_2$O. 
The target array is contained in an AlMg3 safety hull
together with
necessary shielding and cooling tubes, as shown in Fig.~\ref{fig:target}.
The cooling agent is heavy water at a mass flow rate of 22\,kg/s.
The minimal proton beam path length in D$_2$O 
from the target hull to the front Zircaloy tube 
is 14\,cm.
The target assembly is inserted into a vacuum tube 
penetrating the heavy water tank.
This tank has an inside diameter of 160\,cm and a height of 180\,cm and 
contains 3300\,liter of D$_2$O at room temperature
serving as thermal moderator for the primary neutron flux.

%

\subsection{Calculation of the primary neutron yield}
\label{sec:primary-neutrons}

\begin{figure}[htb]
\centering
\includegraphics[width=0.30\textwidth]{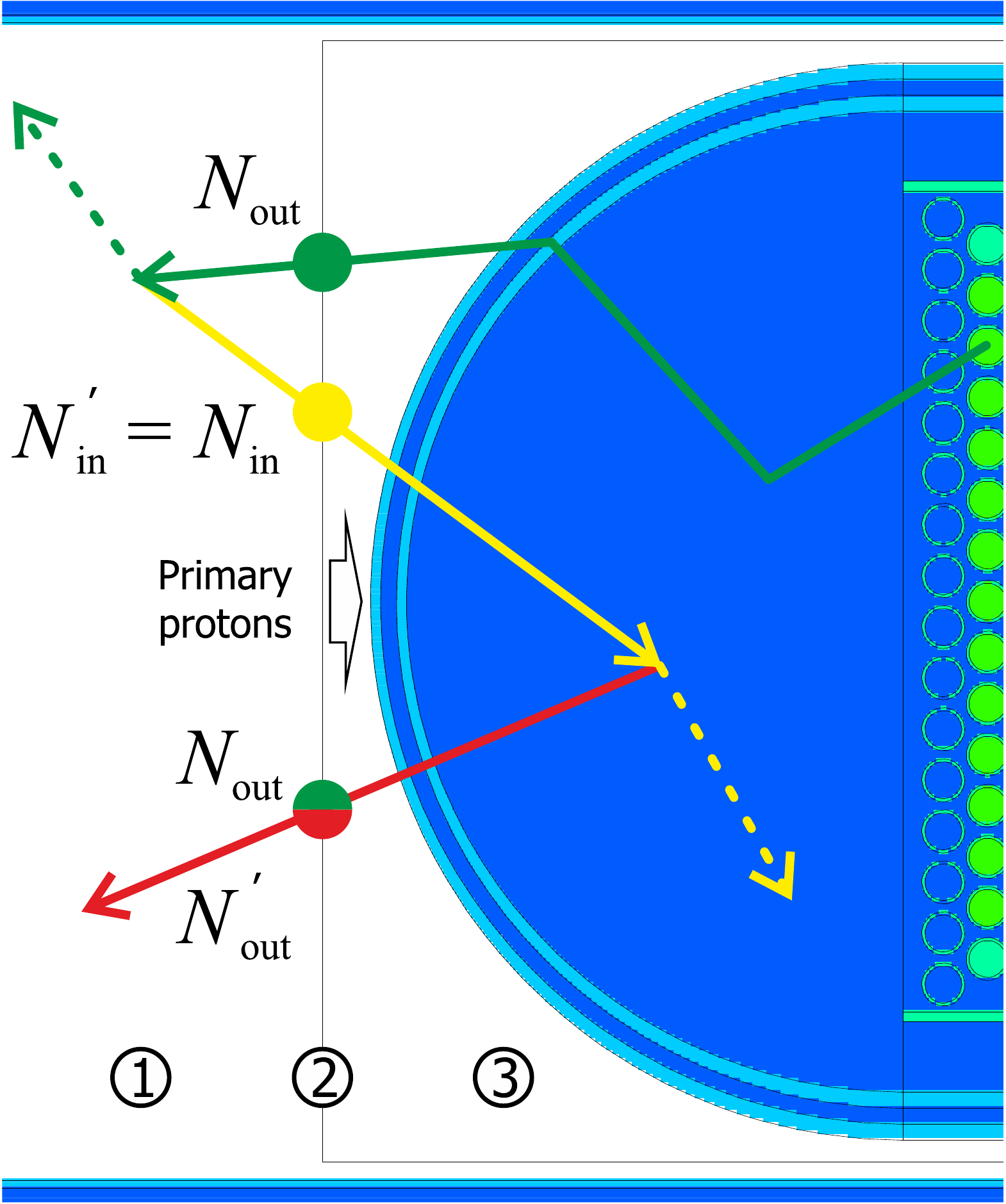}
\caption{Sketch of parts of the MCNPX spallation target geometry model 
showing possible neutron trajectories following 
primary neutron production by beam protons 
modeled within MCNPX.
(1) Tagging region, 
(2) MCNPX target envelope,
(3) MCNPX target cell. 
The cylindrical flow guides are not shown
and irrelevant for the simulation.
}
\label{fig:n-production-sketch}
\end{figure}

A full simulation of the neutron production and moderation in the
UCN source setup was conducted 
in order to establish a consistent description of the activity measurements.
The neutron flux density distribution starting from the proton beam impact
on the spallation target was calculated using
the Monte Carlo radiation transport code MCNPX version 2.7.0 \cite{Pelowitz2011}
with standard S($\alpha,\beta$) tables. 
The MCNPX geometry description followed in detail 
the ``as-built'' construction of the UCN source,
including a realistic simulation model of the spallation
target array, 
cooling arrangement and target container \cite{Wohlmuther2006}.
In the present study the primary neutron yield
from the target was simulated 
as illustrated in Fig.~\ref{fig:n-production-sketch}.
The neutron yield is calculated by surrounding the geometric representation
of the physical target, the target cell, by a target envelope.
The source of primary protons was placed inside this MCNPX target cell, 
but outside the target entrance sphere.
Neutrons produced by nuclear reactions inside the target cell
which leave the target cell will cross the target envelope and 
directly contribute to the outgoing flux 
$N\_{out}$
(green trajectories in Fig.~\ref{fig:n-production-sketch}). 
However, neutrons produced in the target which left
the target cell and returned after being scattered in the 
surrounding structures will be monitored and contribute
to the incoming flux $N_{in}$
and tagged incoming flux $N'_{in}$
(yellow arrow)
to avoid double counting of neutrons.
If such neutrons 
once again leave the target cell, they will be monitored as
well and contribute to the tagged outgoing flux ($N'_{out}$, red arrow).

This model allows us to calculate the
primary neutron yield from the target 
$N_0$ = $N_{out}$ - $N'_{out}$.
$N_0$ is the number of neutrons produced in the target cell by the 
incident proton beam and subsequent particle cascade(s),
excluding those produced by secondary particles 
(or by their products) that 
passed the target envelope 
but then were scattered back 
and passed the target cell once again.

Another way to calculate this quantity is 
to compute it from the data available in the MCNPX output summary tables, 
that can be a
computationally tedious and strictly geometry-dependent procedure (as in \cite{Wohlmuther2006}).
In contrast, the method used here depends only on few surfaces that 
define the target envelope and the cells
that are used for tagging.

The result of our calculation is a neutron yield of 7.27 neutrons per primary proton 
at the target \cite{Talanov2013} which is in good agreement 
with the previous estimate of  
7.62 neutrons per primary proton at the target from \cite{Wohlmuther2006},
within the 5$\%$ estimated error of the calculations.

This value can be compared to the measured neutron yields for the 61\,cm long Pb targets 
as reported in \cite{Bartholomew1966}, which 
are given in Tab.~\ref{Tab_40-02}
after interpolation to the incident proton beam energy of 590\,MeV.
The comparison shows that the design of the spallation target of the UCN source 
provides effective neutronic performance, ensuring at the same time the optimal 
thermo-mechanical behavior required for the high intensity beam conditions.

\begin{table*}[h]
\centering
\begin{tabular}{|c|c|c|c|}\hline 
target & neutron yield       & Pb mass 		& neutron yield \\
       & per proton          & ($\,kg$)  	& per proton   \\
       &                     &        	  & per lead atom       \\
\hline \hline
solid, $\diameter$=10.2\,cm & 9.78 & 57 & 5.9 $\times$ 10$^{-26}$ \\
\hline
solid, $\diameter$=20.4\,cm & 11.2 & 226 & 1.7 $\times$ 10$^{-26}$ \\
\hline
filled ``Cannelloni''       & 7.27 &  93 & 2.7 $\times$ 10$^{-26}$ \\ \hline
\end{tabular}
\caption{
The result of our calculation for the primary neutron yield per incident beam proton
from the filled ``Cannelloni'' spallation target of the UCN source 
compared to the measured neutron yields from solid 61\,cm long Pb targets
with diameters $\diameter$
as reported in \cite{Bartholomew1966}
interpolated to the PSI proton beam energy.
}
\label{Tab_40-02}
\end{table*}

\begin{figure}[htb]
\centering
\includegraphics[width=0.5\textwidth]{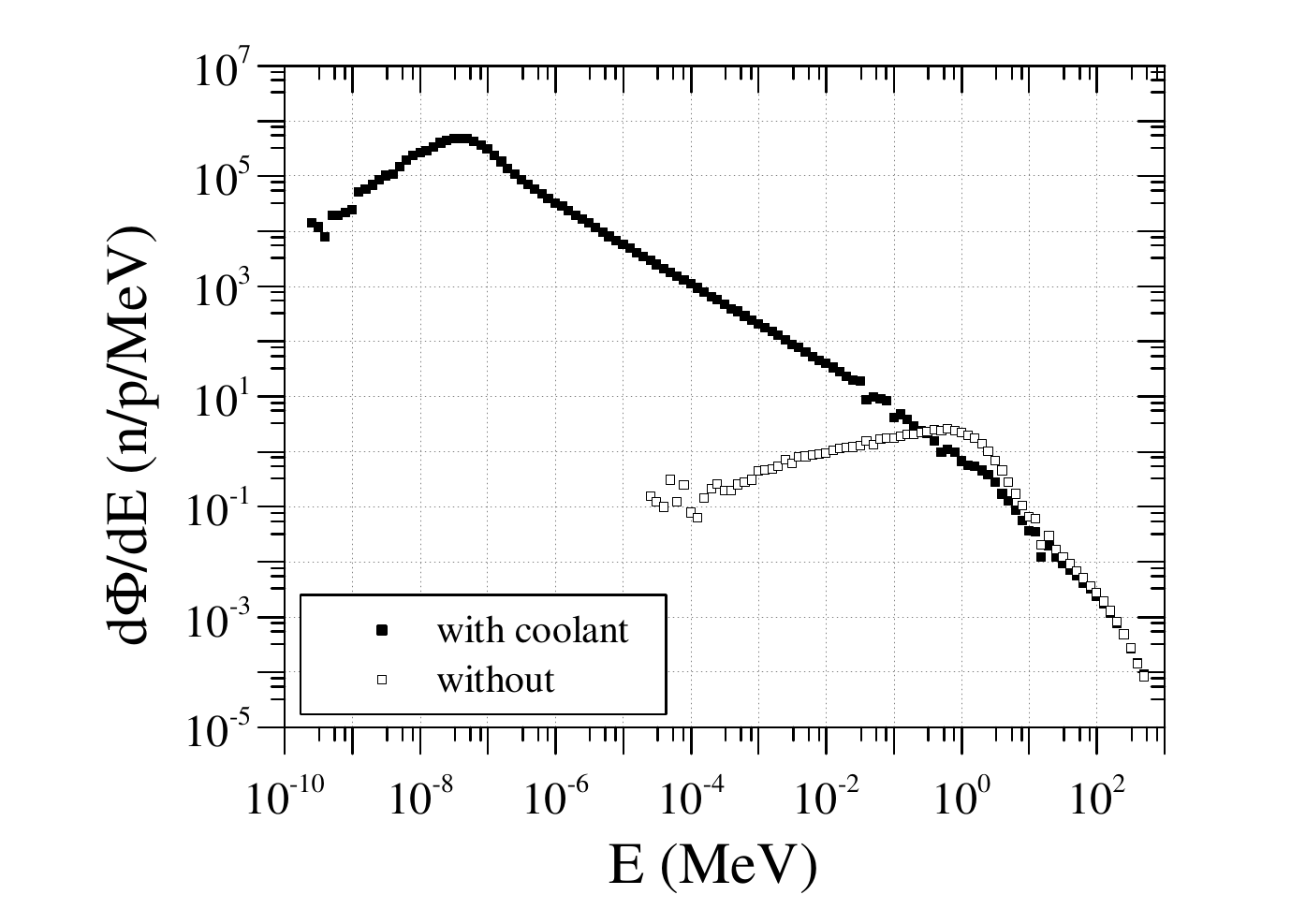}
\caption{
Calculated energy spectrum of the neutrons, given in neutrons per proton per MeV,
at the point of leaving the ''target envelope'' of Fig.\ref{fig:n-production-sketch}
for the cases 
with (filled squares) and without (empty squares) D$_2$O coolant inside the target envelope.}
\label{Fig_10-03}
\end{figure}

Due to the D$_2$O cooling of the spallation target, 
neutrons 
can already be thermalized inside the target safety hull
and either be captured or
leave the target cover pre-moderated.
Fig.~\ref{Fig_10-03} compares the energy spectra 
of neutrons at the point of leaving the target envelope
for the case simulated 
with and without D$_2$O inside the target envelope.
Neutron energies down to the thermal (meV) region demonstrate  
the moderation effect already inside the target envelope.


\section{Gold foil activation measurements}
\label{experiment}


%
The number of produced UCN is directly proportional to the number of 
spallation neutrons and the 
number of thermal neutrons 
entering the solid deuterium converter \cite{Kirch2010}.
Many restrictions, most notably the very high radiation environment,
prevent access to the solid deuterium moderator vessel. 
However, the thermal neutron flux can 
be measured in a distance of about 1\,m from this 
region of interest, 
outside the UCN vacuum and heavy water tank. 
These measurements can directly be compared with 
the numerically estimated neutronic performance 
of the target-moderator system and thus serve to indirectly check the neutron flux 
delivered to the solid deuterium moderator, 
which can be predicted by 
the simulations as well.

\subsection{Measurement principle}

The measurement technique is based on the neutron activation 
of gold foils,
\begin{equation}
\mathrm{
{^{197}Au} + n \rightarrow   {^{198}Au} \rightarrow  {^{198}Hg} + e^{-} + \gamma      . }
\end{equation}

The $\beta$-decay half-life of $^{198}$Au is 2.6947(3)\,days 
with well known $^{198}$Hg
$\gamma$ intensities and energies of 
411.8\,keV
for the main transition, and
675.9\,keV
1087.7\,keV
\cite{ENSDF}.
The initial gold activity is calculated
from the measured absolute $\gamma$ intensity.
This activity can then be compared to the one derived
from a full simulation which takes into account
the proton beam energy and shape,
neutron production via spallation, 
neutron moderation,
neutron transport through the system
and activation of the
$^{197}$Au via neutron capture 
for the neutron flux and energy spectrum at 
the measurement position.

\subsection{Setup for the gold-foil irradiation}

\begin{figure}[htb]
\centering
\includegraphics[width=0.45\textwidth]{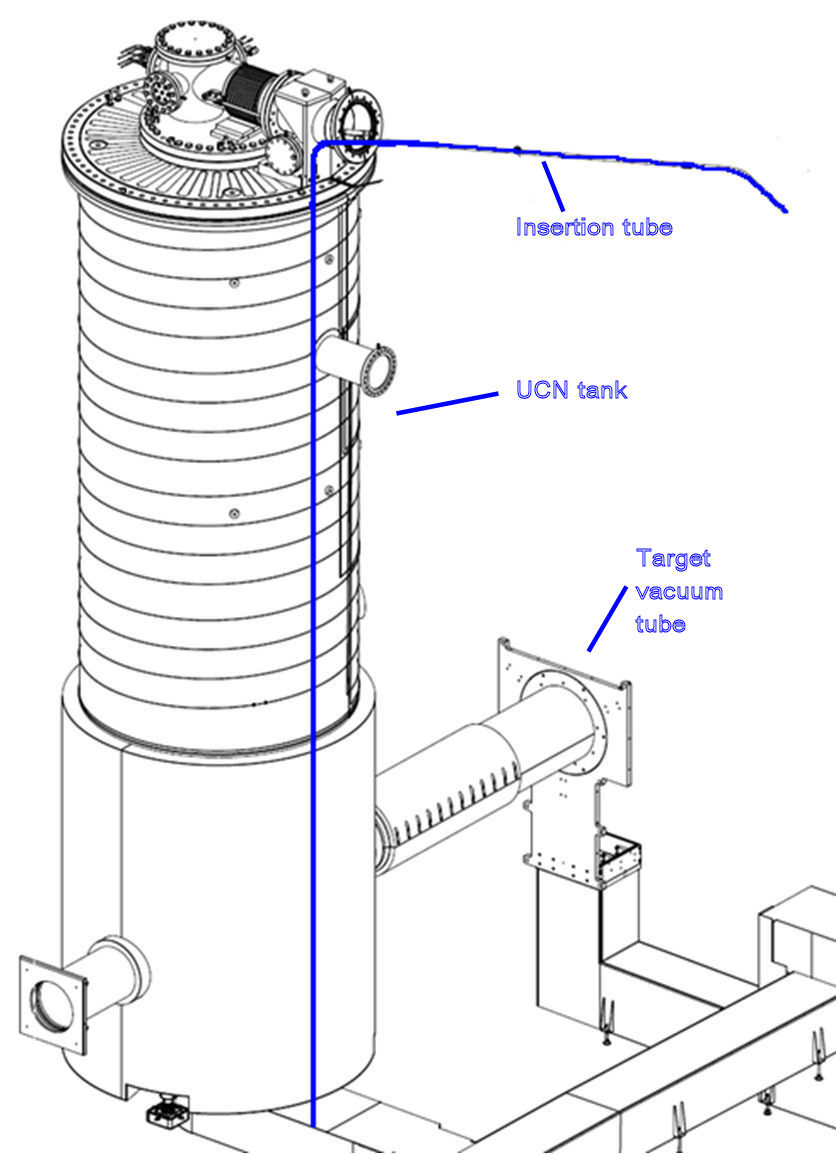}
\caption{Drawing of the 7\,m high UCN tank. 
The vacuum tube for the spallation target penetrates the 
D$_2$O tank which reaches 143\,cm above the center of the target.
The Al ''insertion tube'' along the tank is indicated.
}
\label{ucn-tank}
\end{figure}

The experimental setup uses 
gold foils deployed
on a nylon rope along the UCN tank.
The drawing of the tank in Fig.~\ref{ucn-tank}
shows the position of the aluminum tube which contains the rope. 
Nylon was used to minimize activation of the rope material.
The 16\,mm inside diameter tube horizontally penetrates about 3\,m of the biological 
shielding of the UCN source and then turns down
along the outside of the tank to the floor with a reduced inner 
diameter of 12\,mm. 
There the sharp bend and a welding connection pose the tightest 
constraints for the insertion of any device.
This tube is the best 
access port 
to measure the thermal neutron flux.
It is closed during regular operation.

The length of the insertion tube was measured with a steel rope
pushed down the tube and then extracted again.
The tube length 
of 10.70$\pm$0.01\,m is
in agreement with the design drawing of 10.72$\pm$0.02\,m.
Pushing the steel rope one could feel a resistance after 4.75\,m into the tube
which was identified as welding joint, where the 
inner diameter is reduced to estimated 8.5\,mm.
A similar measurement was done using a nylon rope which was 
compressed and slightly curled inside the tube.
Two measurements yielded 10.82$\pm$0.01\,m
which define a correction factor of 0.989$\pm$0.002 
to the nylon rope length.
To account for eventual changes of compression or twisting
when the rope with the foils was finally inserted,
e.g. due to friction or roughness changes, 
a conservative estimate of the position error of 5\,cm was used in the analysis.
Although the errors on the position of the foils will not be independent
because of the rope, we can neglect this, 
as a better position resolution is not relevant in this study.

\begin{figure}[htb]
\centering
\subfigure{\includegraphics[width=0.2\textwidth]{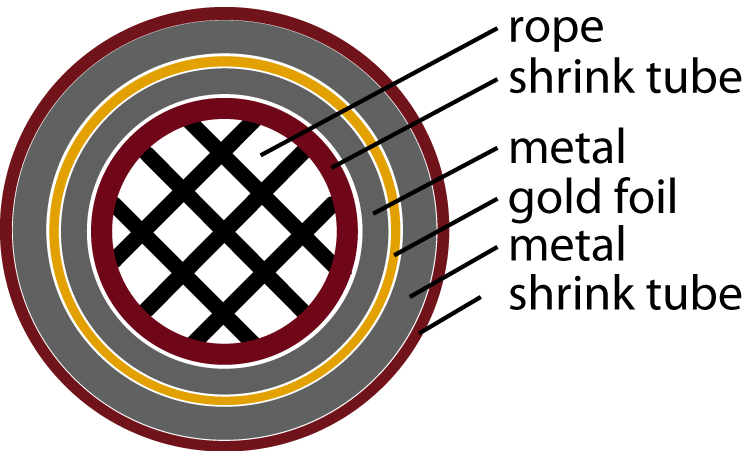}}
\subfigure{\includegraphics[width=0.2\textwidth]{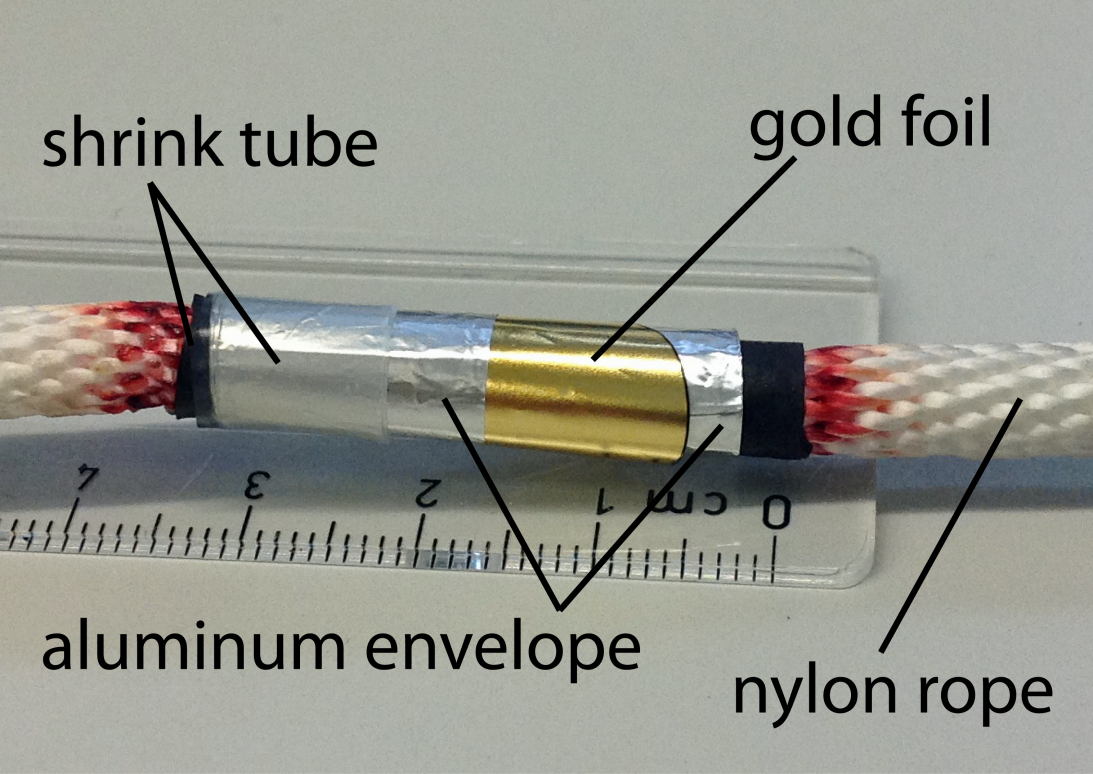}}
\caption{
a) Sketch of the mounting assembly for the gold foils on the nylon rope.
After a layer of heat-shrink tubing the aluminum-foil-covered 
gold foil was wrapped around the rope. 
Then another shrink tube held the sandwich in place. 
In the measurement with Cd the additional foil covered
both sides of the gold. 
The heat-shrink tubing was used to secure the foils in place.
\newline
b) Photo of the rope during probe assembly with part of the Al foil on the outside removed.}
\label{gold-foil-mount}
\end{figure}

The foils used were 
99.95$\%$ purity gold ($^{197}$Au) with a thickness of 25\,$\mu$m from Goodfellow,
laser-cut into discs with 25\,mm diameter.
After cleaning with isopropyl alcohol, the foils were weighed three times with a
laboratory balance 770-12 by Kern und Sohn.
The averaged results are listed in 
tables \ref{tab:result-cd} and \ref{tab:result-h} with a 1\,mg error on the
calibration of the balance.
Finally, the foils were attached to the rope using heat-shrink tubing and 
aluminum foils
as shown in Fig.~\ref{gold-foil-mount}.

Three measurement assemblies were used, 
one for the test measurement, 
one for the height profile,
and one with a cadmium shielded foil.
The thicknesses of the used materials were: 
rope = 6.7\,mm,
aluminum foil = 0.01\,mm, 
heat-shrink tubing = 0.4\,mm.
In the setup with the additional 0.55\,mm thick Cd covers 
the rope thickness was reduced to 2.2\,mm.
The Cd covered foil was positioned 
in the center of two uncovered gold foils 19\,cm apart.
The intended position of the gold foils were accurately
marked on the rope before assembly in order to guarantee
the position. The foil positions are listed in
tables \ref{tab:result-cd} and \ref{tab:result-h}.

\subsection{Irradiation of the samples at the UCN source}

The following irradiations were conducted using three setups:
\begin{enumerate}

\item{Preparatory measurement with a single gold foil positioned 
at beam height
to test the feasibility of the foil insertion and irradiation.
2012-10-18: proton pulse length = (57\,$\pm$~4)\,ms;
beam current = (2186\,$\pm$~44)\,mA.
}

\item{A neutron-energy-sensitive measurement with two gold foils and 
one additional Cd covered gold foil.
2012-11-14:
proton pulse length = (507\,$\pm$~4)\,ms;
beam current = (2195\,$\pm$~44)\,mA.
}

\item{A height profile measurement along the UCN tank with 
16 gold foils positioned from about
1\,m below to 5\,m above the beam plane.
2012-11-21:
proton pulse length = (2007\,$\pm$~4)\,ms;
beam current = (2197$\pm$~44)\,mA.
}

\end{enumerate}

The error on the proton beam current is estimated to be about 2\,$\%$ 
resulting from the absolute calibration error of the 
proton beam current monitor (MHC1),
installed on the proton beam line after the exit of the cyclotron.
The given beam pulse length
reflects the sum of the pilot and main beam pulse. 
The maximum error on the irradiation time of $\pm$~4\,ms is estimated
from the 1\,ms rise-and-fall time of the beam kicker 
power supply \cite{kicker2005} adding up pilot and main pulse transients.

\subsection{Determination of the gold activation via $\gamma$-ray analysis}

The $\gamma$-ray analysis following the activation was performed at
the radioanalytical laboratory of PSI 
which is an official Swiss accredited 
laboratory for the survey and analysis of radioisotopes 
for emission and incorporation.
In our measurements we employed 
an efficiency calibrated 
high purity N-type Ge detector from ORTEC
with 101~cm$^3$ active volume 
to determine the $\gamma$ intensity.

\begin{figure}[htb]
\centering
\includegraphics[width=0.5\textwidth]{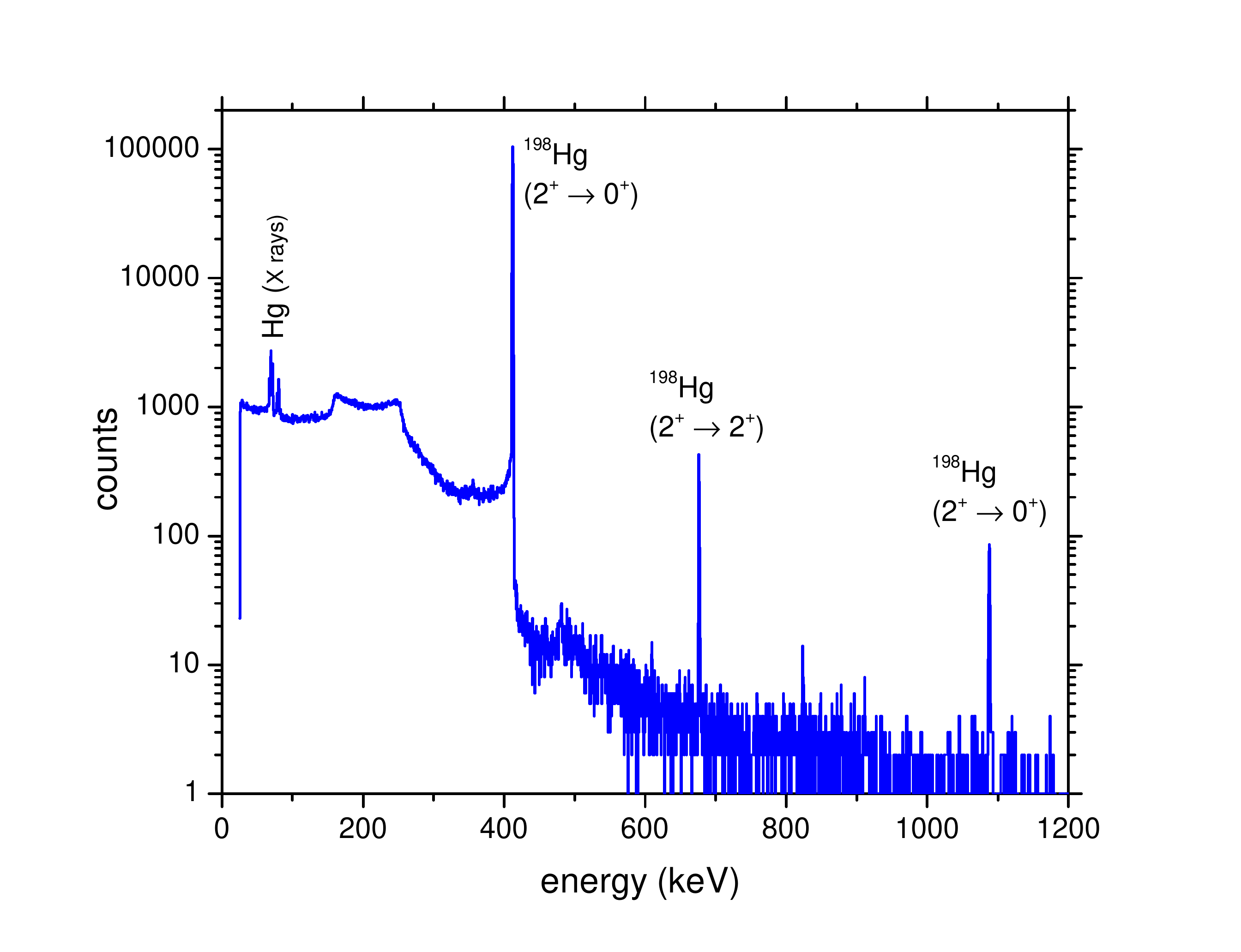}
\caption{
Energy spectrum of $\gamma$-rays observed for foil 4 
after 20 minutes
of measurement time.
}
\label{fig:gamma-spectrum}
\end{figure}

The 25\,mm diameter
gold foil shape was selected to match the requirements for the 
geometry setup
previously calibrated to $\pm$5$\%$ detection efficiency.
The gold foil was pressed to lie flat
on the measurement position before the measurement.
Measurements lasted between 10\,min and 40\,min,
depending on the sample activity.
In Fig.~\ref{fig:gamma-spectrum}
we show the measured $\gamma$ energy spectrum for foil 4.
The dominant line at 411.8\,keV 
is from the $^{198}$Hg ~$2^+\rightarrow 0^+$ transition.
The electronic K transition X rays in Hg are visible at 70 and 80\,keV.

The spectroscopy program InterWinner~5.0 by ORTEC
was used for the energy spectrum analysis.
The measured intensity is corrected for 
the time period between $\gamma$ measurement and irradiation, 
recorded with 1\,s accuracy, DAQ dead time and energy dependent detector efficiency.
The measured environmental background on the 10$^{-4}$ Hz/channel 
level is automatically subtracted but negligible here.
The background under the relevant $\gamma$ peak is due to 
Compton scattering of higher-energy $\gamma$ rays and automatically subtracted
by the software.
The results of our analysis for the measured activities of the
foils and their 
positions are summarized in 
tables \ref{tab:result-cd} and \ref{tab:result-h}.
The $\pm$5$\%$ error on the absolute efficiency calibration 
of the $\gamma$ detector is included in the
final error.
A comparison with the simulation 
can be done after normalization 
with irradiation time (beam pulse length) and proton beam current.

The results of Tab.~\ref{tab:result-cd} and \ref{tab:result-h}
plotted in 
Fig.~\ref{results-comparison} show that 
the activity maximum corresponds to the beam plane position.
The drop in activity
at about 1.4\,m, corresponds
to the top of the heavy water tank.

\begin{table*}[htb]
\begin{center}
    \begin{tabular}{ | c | c | c  | c | c |}
    \hline
    measurement        &     position      &       foil mass       & specific activity     \\ 	
                       &       ($cm$)      &         ($mg$)        &  ($kBq$/g)            \\ \hline\hline
    preparatory        &   0.0   & 174.9 $\pm$ 1.0       & 1378.  $\pm$ 122.     \\ \hline\hline
    Au-down            & 107.9   & 262.5 $\pm$ 1.0       &  169.5 $\pm$ 9.3       \\ \hline
		Au-Cd-center       & 117.4   & 261.4 $\pm$ 1.0       &   22.8 $\pm$ 1.2        \\ \hline
		Au-up              & 126.9   & 251.9 $\pm$ 1.0       &  135.0 $\pm$ 7.3        \\ \hline 
\end{tabular}
\caption{Results of the preparatory measurement (single foil) and the measurement with the 
Cd-covered foil vertically centered between two gold foils 19\,cm apart.
The position is given as vertical distance to the proton beam plane =~0, 
$+$ is in upward direction. 
The absolute position uncertainty for every measurement point is estimated 
to be $\pm$5\,cm, but this uncertainty is not relevant for this measurement. 
The measured specific activity is given by the measured $\gamma$ activity 
corrected for time period between irradiation and measurement,
decay branching, DAQ dead time and detector efficiency,
normalized to an irradiation time of 1\,s at a proton beam current of 2.2\,mA.
The given uncertainty includes the $\pm$5$\%$ calibration error 
on the absolute efficiency of the
$\gamma$ detector. 
}
\label{tab:result-cd}
\end{center}
\end{table*}

\begin{table*}[htb]
\begin{center}
    \begin{tabular}{ | c | c | c | c | c | c |}
    \hline
    foil number        &     position      &       foil mass       & specific activity    \\ 	
                       &       ($cm$)       &         ($mg$)        &  ($kBq$/g)           \\ \hline\hline
    1  	&  -73.3    & 233.4 $\pm$ 1.0  	 &  388.   $\pm$ 21.		 \\ \hline
    2  	&  -48.6    & 232.2 $\pm$ 1.0  	 &  753.   $\pm$ 41.		 \\ \hline
    3  	&  -23.9    & 241.0 $\pm$ 1.0  	 & 1198.   $\pm$ 65.		 \\ \hline
    4  	&    0.9    & 237.0 $\pm$ 1.0  	 & 1333.   $\pm$ 72.		 \\ \hline
    5  	&   25.6    & 242.9 $\pm$ 1.0  	 & 1040.   $\pm$ 56.		 \\ \hline
    6  	&   50.3    & 227.7 $\pm$ 1.0  	 &  654.   $\pm$ 35.		 \\ \hline
    7  	&   75.0    & 242.3 $\pm$ 1.0  	 &  359.   $\pm$ 19.		 \\ \hline
    8  	&   99.8    & 230.0 $\pm$ 1.0 	 &  191.   $\pm$ 10.		 \\ \hline
    9  	&  124.5    & 237.0 $\pm$ 1.0  	 &  147.   $\pm$  8.	   \\ \hline
    10 	&  149.2    & 245.0 $\pm$ 1.0  	 &   68.2  $\pm$  3.7		 \\ \hline
    11 	&  173.9    & 242.1 $\pm$ 1.0  	 &   50.7  $\pm$  2.7		  \\ \hline
    12 	&  198.6    & 211.0 $\pm$ 1.0  	 &   45.0  $\pm$  2.5		 \\ \hline
    13 	&  248.1    & 224.8 $\pm$ 1.0  	 &   35.3  $\pm$  1.9		  \\ \hline
    14 	&  297.5    & 226.6 $\pm$ 1.0  	 &   26.3  $\pm$  1.4		  \\ \hline
    15 	&  396.4    & 234.4 $\pm$ 1.0    &   14.5  $\pm$  0.8		  \\ \hline
    16 	&  495.3    & 241.6 $\pm$ 1.0 	 &    1.45 $\pm$  0.10		\\ \hline
\end{tabular}
\caption{Results of the height profile measurement.
Positions and uncertainties as described in the caption of Tab.~\ref{tab:result-cd}.
}
\label{tab:result-h}
\end{center}
\end{table*}


\section{Numerical simulation of the thermal neutron flux}
\label{simulation}

\begin{figure}[h]
\centering
\includegraphics[width=0.45\textwidth]{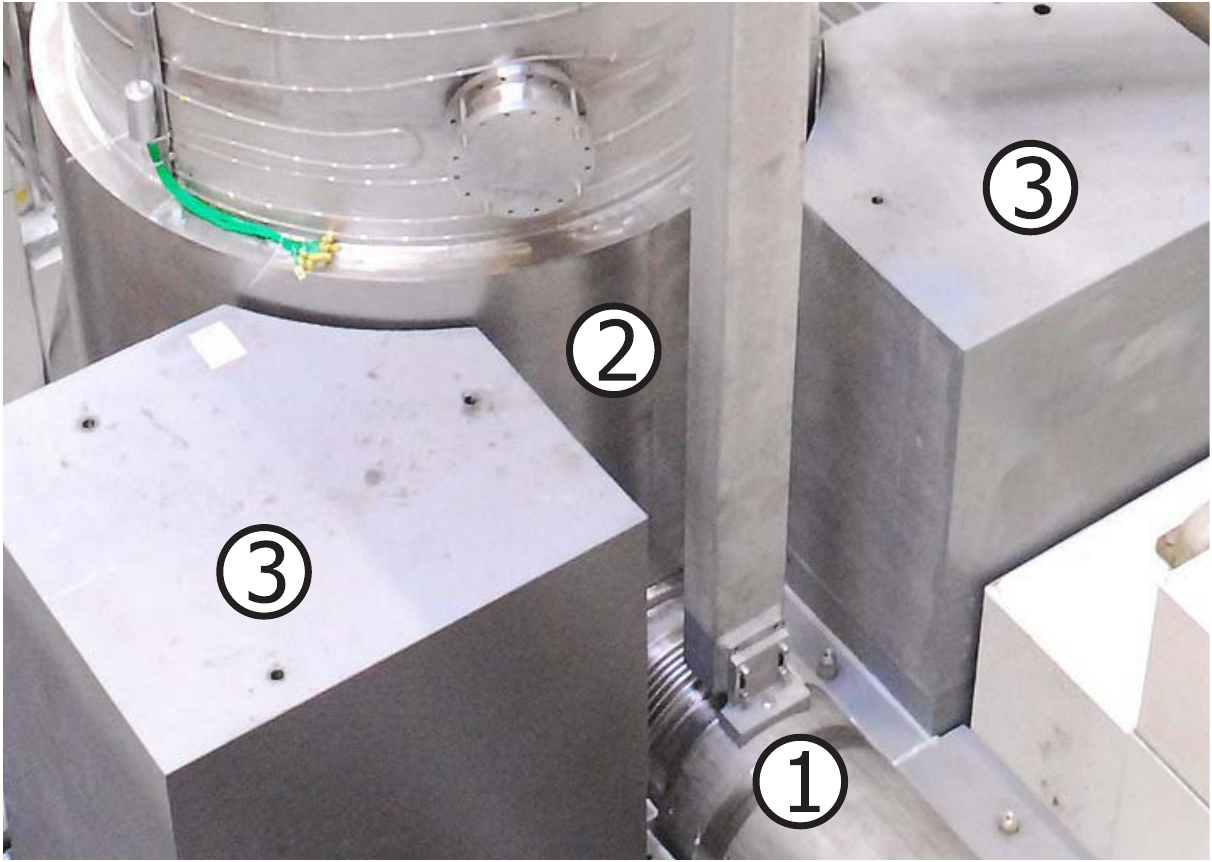}
\caption{Photo of the bottom part of the UCN tank during construction of the UCN source:
(1) Outer vacuum vessel for the spallation target, 
(2) D$_2$O tank surrounded by (3) iron shielding.
}
\label{fig:target_and_concrete}
\end{figure}
\begin{figure}[htb]
\centering
\includegraphics[width=0.4\textwidth]{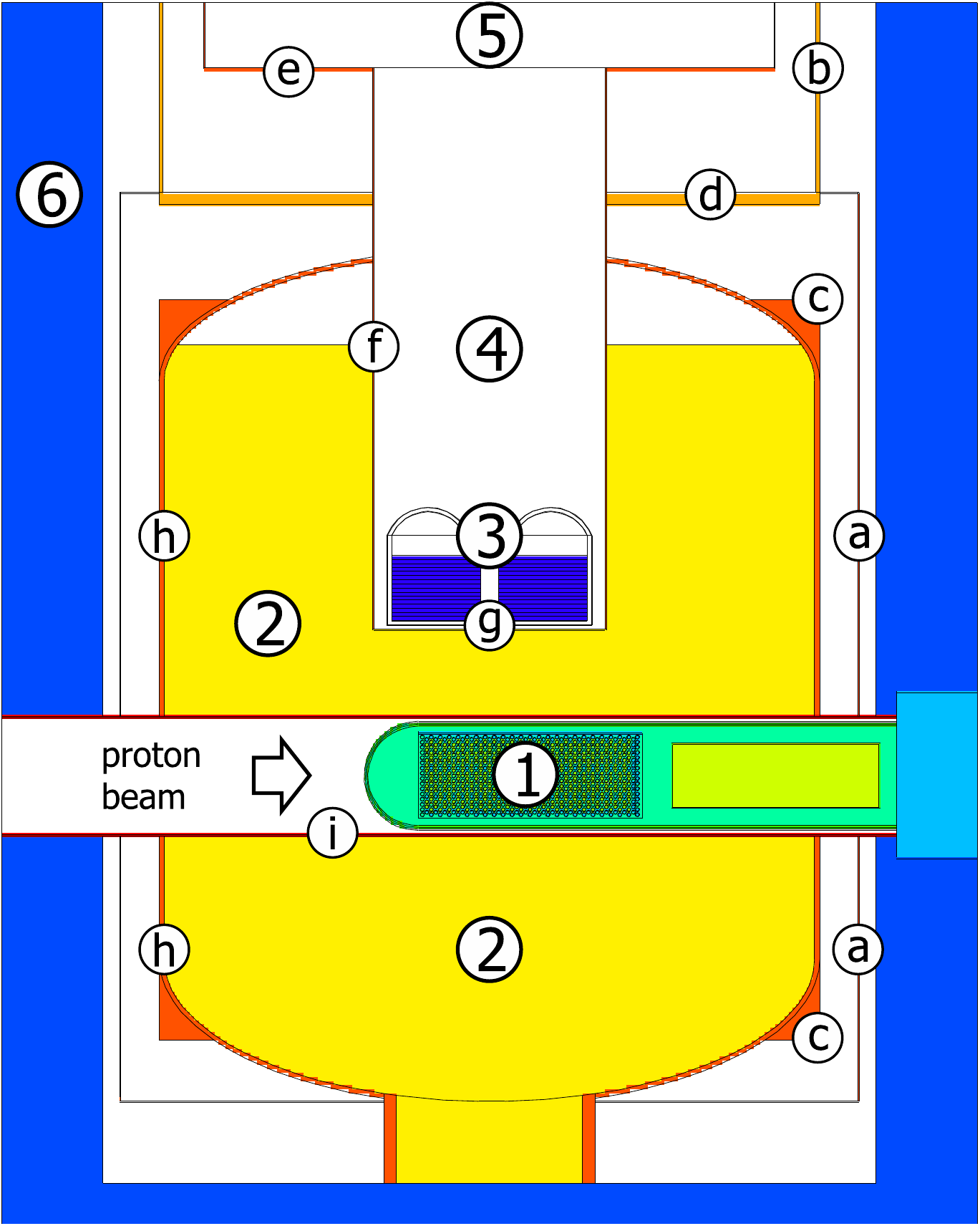}
\caption{Drawing of the MCNPX geometry model of the central part of the UCN source: 
(1) ``Cannelloni'' spallation target (Zr,Pb) with the forward neutron lead shielding (right side), 
(2) D$_2$O tank (Al), 
(3) solid deuterium vessel (AlMg3, AlMg4.5, Al), 
(4) vertical UCN guide (Al), 
(5) UCN storage vessel and shutter (Al), 
(6) innermost shielding, and structural details of the UCN source a) --- i) (see Table~\ref{tab:source-sim-par}).}
\label{Fig_30-01}
\end{figure}

%
In the present simulation we have used a 
description with --- in comparison to the early design \cite{Atchison2002} --- 
many more construction details
of the UCN tank and 
the area surrounding the target,
such as the support structure of the 
heavy water moderator tank, the outer shell of the UCN source,
UCN storage vessel etc. \cite{Talanov2013a}. 
A vertical cut of the MCNPX geometry model with definitions used in the particle transport simulation 
is shown in Fig.~\ref{Fig_30-01},
some important dimensions
are given in Tab.~\ref{tab:source-sim-par}.
A photo of the heavy water (bottom) part of the UCN tank during 
construction displaying 
some details of the surroundings
is shown in Fig.~\ref{fig:target_and_concrete}.

\begin{table*}[h]
\centering
\begin{tabular}{| c | c | c | c |}
\hline
& part  & material &  thickness ($cm$)  \\ \hline \hline
a) & wall of the UCN tank (bottom)    		& Al		& 0.3	  \\	\hline 
b) & wall of the UCN tank (top)        		& steel	& 1.0	  \\	\hline 
c) & bottom and top ring of the D$_2$O tank       &	Al	    &	20		\\	\hline 
d) & top lid of the D$_2$O tank 	 				&	steel		&	3.0		\\	\hline 
e) & wall of the UCN storage vessel       & Al 	    & 0.3		\\	\hline 
f) & vertical UCN guide wall		 					&	Al			& 0.3		\\	\hline 
g) & vertical UCN guide bottom						&	Al			& 0.3		\\	\hline 
h) & wall of the D$_2$O tank		        	&	Al			&	1.2		\\	\hline 
i) & wall of proton beam tube   					&  Al     &  0.25   \\ \hline 
\end{tabular}
\caption{
List of material thicknesses of important structural parts 
(see Fig.~\ref{Fig_30-01}) of the MCNPX geometry model.
}
\label{tab:source-sim-par}
\end{table*}

\begin{figure}[htb]
\centering
\includegraphics[width=0.5\textwidth]{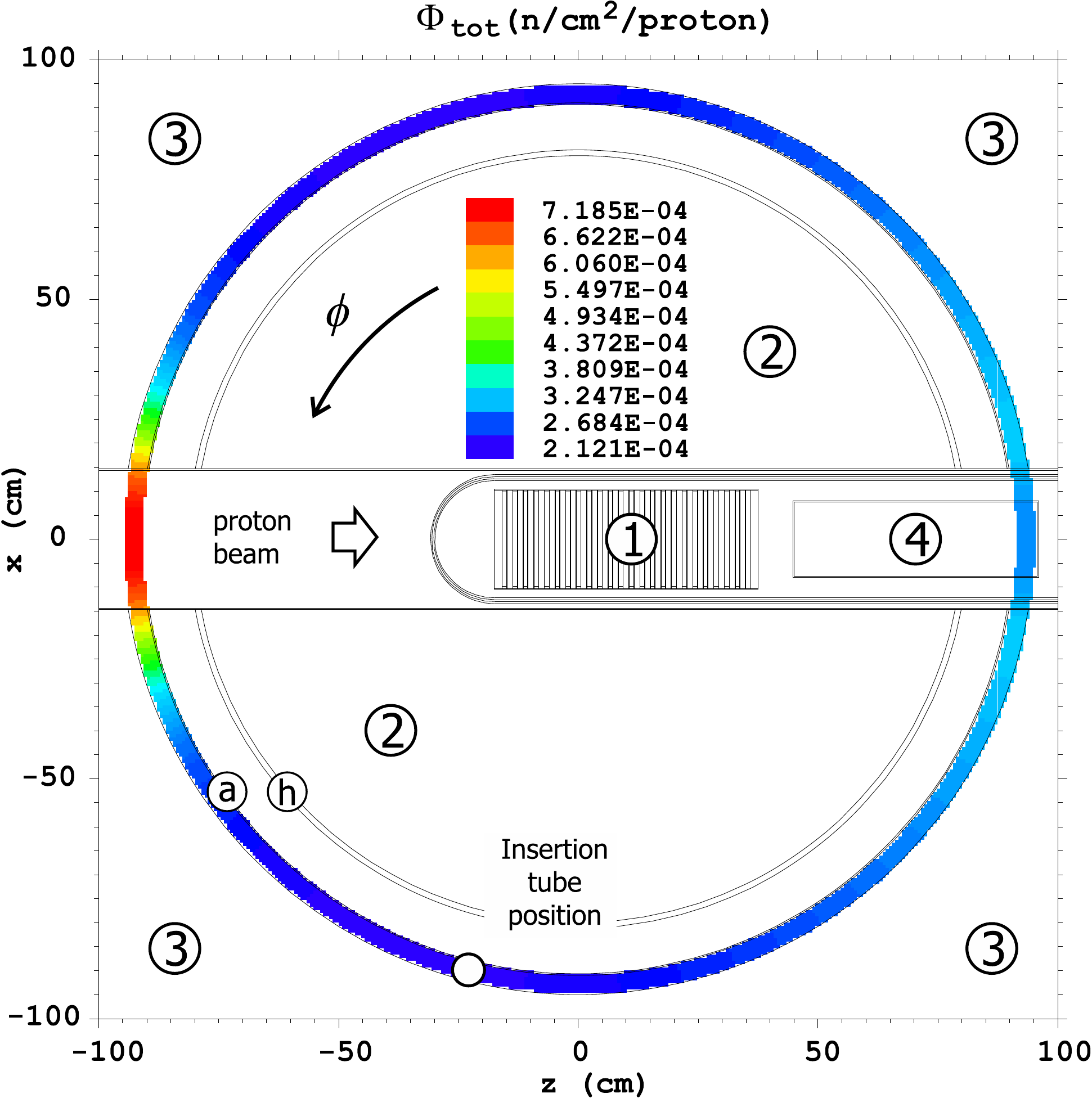}
\caption{Total neutron flux density distribution $\Phi_{tot}$(n/cm$^2/p$)
simulated around the UCN tank 
in the beam plane
versus azimuthal angle $\phi$: 
(1) ``Cannelloni'' spallation target (Zr,Pb) with the (4) forward lead shielding, 
(2) UCN tank, 
(3) innermost shielding (structural details a) and h) defined in Table~\ref{tab:source-sim-par})).}
\label{Fig_30-04}
\end{figure}

\begin{figure}[htb]
\centering
\includegraphics[width=0.5\textwidth]{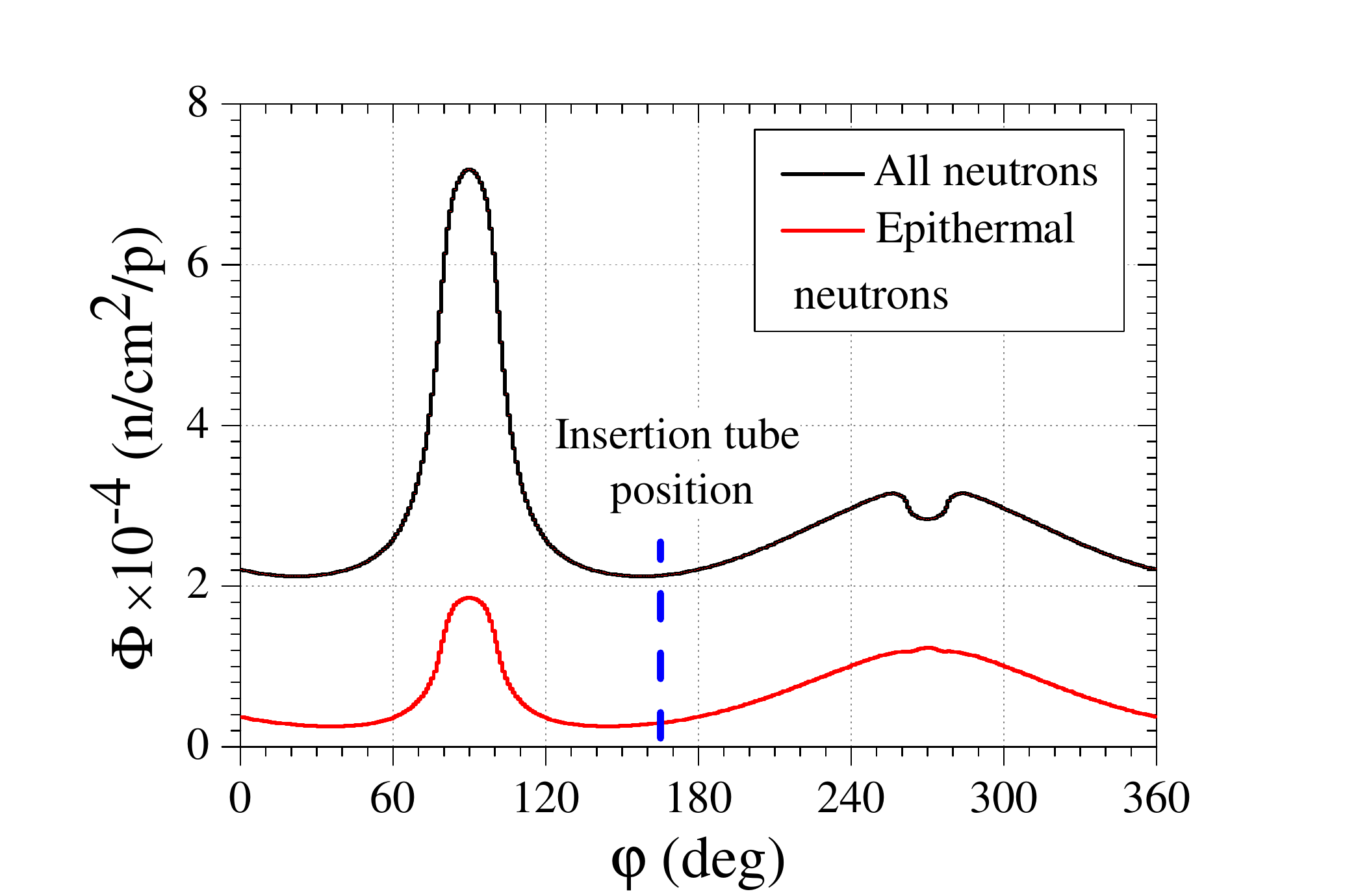}
\caption{
Simulated neutron flux density distribution $\Phi$ per incident proton on the target
around the UCN tank 
in the beam plane 
as a function of the azimuthal angle, for all and for epithermal neutron energies. 
Position $\phi$~=~0 is perpendicular to the direction 
of the proton beam.
The position of the insertion tube for the Au foils 
at about 165$^{\circ}$ is indicated.
90$^{\circ}$ is opposite to the proton beam direction.
}
\label{Fig_30-02}
\end{figure}

After simulation of the initial neutron production,
as described in Sec.~\ref{sec:primary-neutrons},
MCNPX was used to further track the neutrons.
The neutron flux density was simulated in the gap 
between the outer wall of the UCN tank and the innermost 
shielding blocks of the UCN source (see Fig.~\ref{Fig_30-04}).
Fig.~\ref{Fig_30-02}
shows the distribution in the horizontal proton beam plane defined 
through the horizontal axis of the spallation target perpendicular 
to the plane of Fig.~\ref{Fig_30-01}. 
A neutron energy of 0.5\,eV was used as 
lower threshold definition for the epithermal flux component. 
The numerical activity results with and without Cd shielding
however used the full neutron energy information.

The numerical values of the total and epithermal neutron flux density 
from Fig.~\ref{Fig_30-02} for four angles in the proton beam plane are given 
in Tab.~\ref{Tab_30-01}. 
The neutron flux density in the beam plane has a clear maximum  
towards the incoming proton beam ($\phi = 90^{\circ}$)
due to the small amount of material in this region.
The ratio of epithermal to total neutron flux density at this angle 
is $\sim1/3$ larger than at $\phi = 0^{\circ}$ or $180^{\circ}$. 
The dip 
in the total neutron flux density 
at $\phi = 270^{\circ}$
corresponds to the position of the 
lead shielding cylinder 
that is located inside the target downstream of the Zr-Pb array. 
There half of the escaping neutrons possess energies above 0.5\,eV. 
Both, the total and the epithermal flux densities
at $\phi = 0^{\circ}$ are equal to the respective values at $180^{\circ}$
within the statistical error of the simulation, 
as the MCNPX model above the spallation target is axially symmetric.
The fraction of epithermal neutrons 
at the position of the insertion tube 
(Fig.~\ref{Fig_30-02} dashed line)
was calculated to be about 14$\%$.
The variation of both total and epithermal neutron flux densities 
over the azimuthal angle is found insignificant in this position.
The estimated angular uncertainty of $\sim$1.5$^{\circ}$ on the verticality of the insertion-tube
positioning has a negligible effect 
on the results of the simulation.

\begin{table}[h]
\centering
\begin{tabular}{|c|cccc|}\hline
$\Phi$ $\smallsetminus$ angles & $0^{\circ}$ & $90^{\circ}$ & $180^{\circ}$ & $270^{\circ}$ \\ 
\hline \hline
$\Phi_{tot}$              & 2.20 & 7.18 & 2.21 & 2.83 \\ \hline
$\Phi_{epi}$              & 0.37 & 1.85 & 0.38 & 1.23 \\ \hline
$\Phi_{epi}$/$\Phi_{tot}$ & 0.17 & 0.26 & 0.17 & 0.44 \\  
\hline
\end{tabular}
\caption{
Total, $\Phi_{tot}$, and epithermal, $\Phi_{epi}$, neutron flux density 
per incident proton on  the target (10$^{-4}$ n/cm$^2$/p) 
and epithermal fraction 
at different 
angles in the proton beam plane.}
\label{Tab_30-01}
\end{table}

In the MCNPX model 
118 identical gold-foil units were positioned
inside the insertion tube in 5\,cm steps from 
$-$92.5\,cm below to $+$497.5\,cm above the beam plane. 
The geometry model of each gold foil unit 
followed the experimental setup described in Sec.~\ref{experiment}. 
The neutron flux density and energy spectra were simulated
in the exact MCNPX volume of the gold foil of each gold foil unit.

\begin{figure}[htb]
\centering
\includegraphics[width=0.5\textwidth]{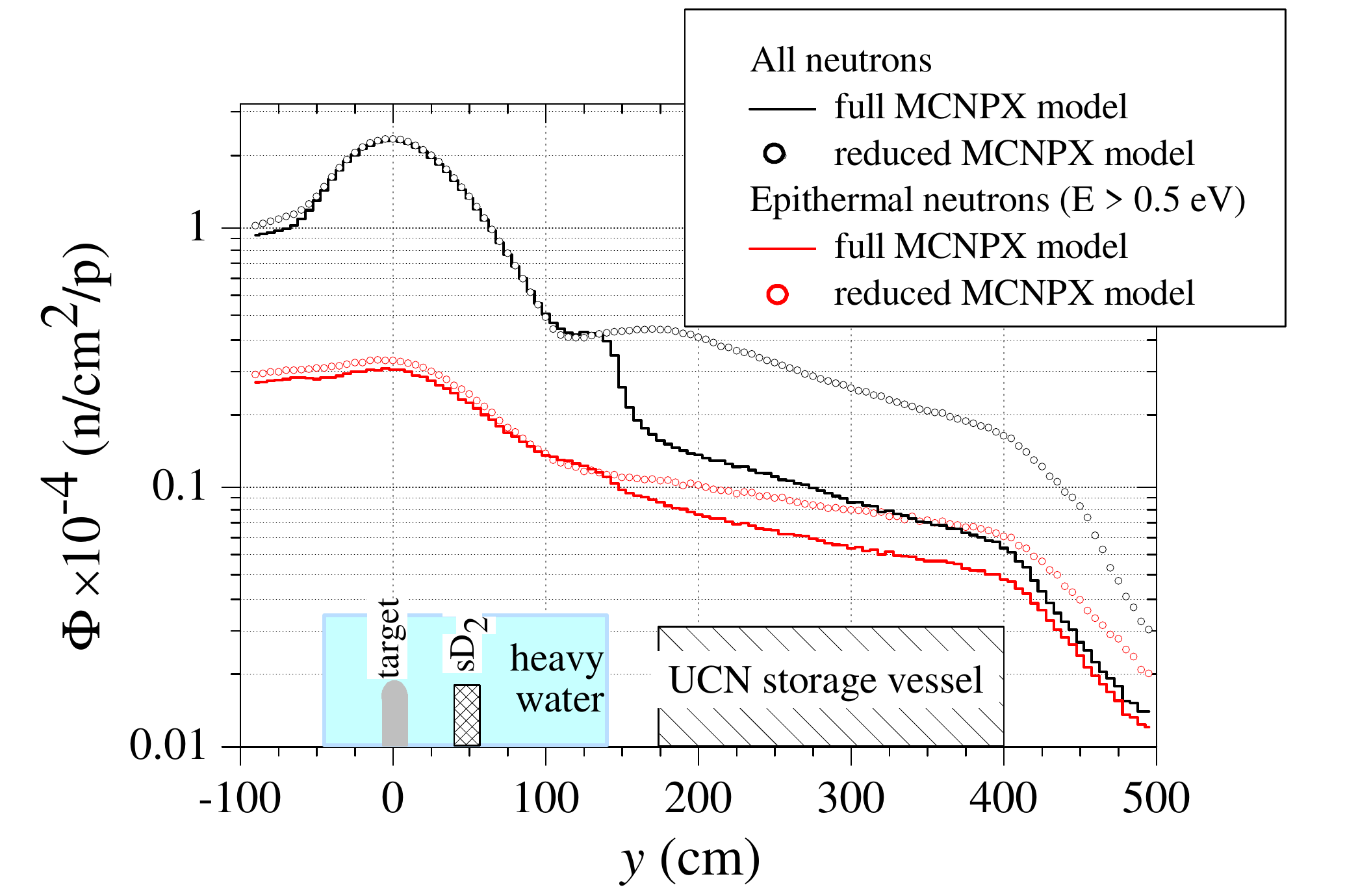}
\caption{
Simulated neutron flux density distribution at the position of the insertion tube
$\Phi$ per incident proton on the target
versus height (y) along the UCN tank
at the position of the insertion tube, 
for all and for epithermal neutron energies. 
Results are compared for the simulations with the full MCNPX model (solid lines), and 
the reduced MCNPX model (open symbols).
Major parts inside the UCN tank are indicated at their approximate y-position.
}
\label{Fig_30-03}
\end{figure}

The material budget and the implemented geometry details 
in the full MCNPX geometry model 
shown in Fig.~\ref{Fig_30-01} contain many details of the setup,
but could in principle still be further improved in terms of minute details.
To evaluate the impact of geometrical uncertainties on the numerical results, 
the MCNPX simulation 
was run 
with a reduced geometry model of the UCN source, 
in which the structural details (a) --- (d) from Fig.~\ref{Fig_30-01} 
were removed.
Neutron flux density distributions
simulated with full and reduced MCNPX
geometry representation are compared in Fig.~\ref{Fig_30-03}.
The statistical error of the simulation 
at beam height (0\,cm)
is below $\pm$2$\%$ 
for both total and epithermal neutron flux.

An increase of the epithermal neutron flux by $\sim$\,10 \% is observed in the simulation with
the reduced MCNPX model between $-$50 and $+$100\,cm height.
At the same time practically no difference in
total neutron flux 
is found in this region.
Above 150\,cm,
the two estimates
differ by more than a factor of two, 
indicating that in the top region more detailed modeling would be required.

A separate simulation study changing the operating temperature and hence the density
of the heavy water moderator 
from 31$^{\circ}$C by $\pm$10$^{\circ}$C 
showed a negligible change of the neutron flux.


\section{Calculation of activities and comparison to measurement results}
\label{results}


The neutron flux density and energy spectra simulated as described in Sec.~\ref{simulation}
were used to calculate the induced $^{198}$Au activity in the gold foils using
the build-up and decay code FISPACT07 \cite{Forrest2007}. 
An automated procedure of activity calculation using the results of the 
MCNPX simulation was provided by the Activation Script \cite{Gallmeier2007}.
The result of the calculation, 
the position dependent specific $^{198}$Au activity 
of the gold foils along the UCN tank,
is given in Tab.~\ref{tab:simulation-result}.

\begin{table}[htb]
\begin{center}
    \begin{tabular}{ | c | c | c |}
    \hline
        position         & A$_{sp}$    &  D$_{MS}$            \\ 
          ($cm$)          &  ($kBq$/\,g)             &      \\ \hline \hline  
      	-73.3    &   438.  	&  -0.12	   \\ \hline
        -48.6    &   599.  	&	  0.23    \\ \hline
      	-23.9    &   944.  	&   0.24	   \\ \hline
       	  0.9    &   1090.  	&   0.20 	 \\ \hline
      	25.6     &   930.  	&   0.11	   \\ \hline
      	50.3     &    626.  	&	  0.04     \\ \hline
        75.0     &    378.   &		-0.05    \\ \hline
        99.8     &    215.   &	 -0.12    \\ \hline
        124.5    &    187.   &  -0.24    \\ \hline
    	  149.2    &     98.  	&	 -0.35    \\ \hline
    	  173.9    &     59.  	&	 -0.15     \\ \hline
     	  198.6    &     49.  	&	  -0.09     \\ \hline
     	  248.1    &     44.  	&	 -0.21     \\ \hline
     	  297.5    &     29.   &	 -0.12	    \\ \hline
     	  396.4    &     23.  &	 -0.48     \\ \hline
     	  495.3    &     3.8  	&  -0.90 	  \\ \hline 
\end{tabular}
\caption{
Results for the height dependent activity simulation.
The position of the gold foil is given as in Tab.~\ref{tab:result-h}.
Simulated specific activities A$_{sp}$
are normalized to an irradiation time of 1\,s at 2.2\,mA proton beam current.
Column~3 gives the relative 
difference between simulation and measurement 
(Tab.~\ref{tab:result-h}) 
D$_{MS}$~=~
(A$_{sp}$(measurement)-A$_{sp}$(simulation))/
(A$_{sp}$(measurement)+A$_{sp}$(simulation)/2.)
}
\label{tab:simulation-result}
\end{center}
\end{table}

\begin{figure*}[h]
\centering
\includegraphics[width=1.0\textwidth]{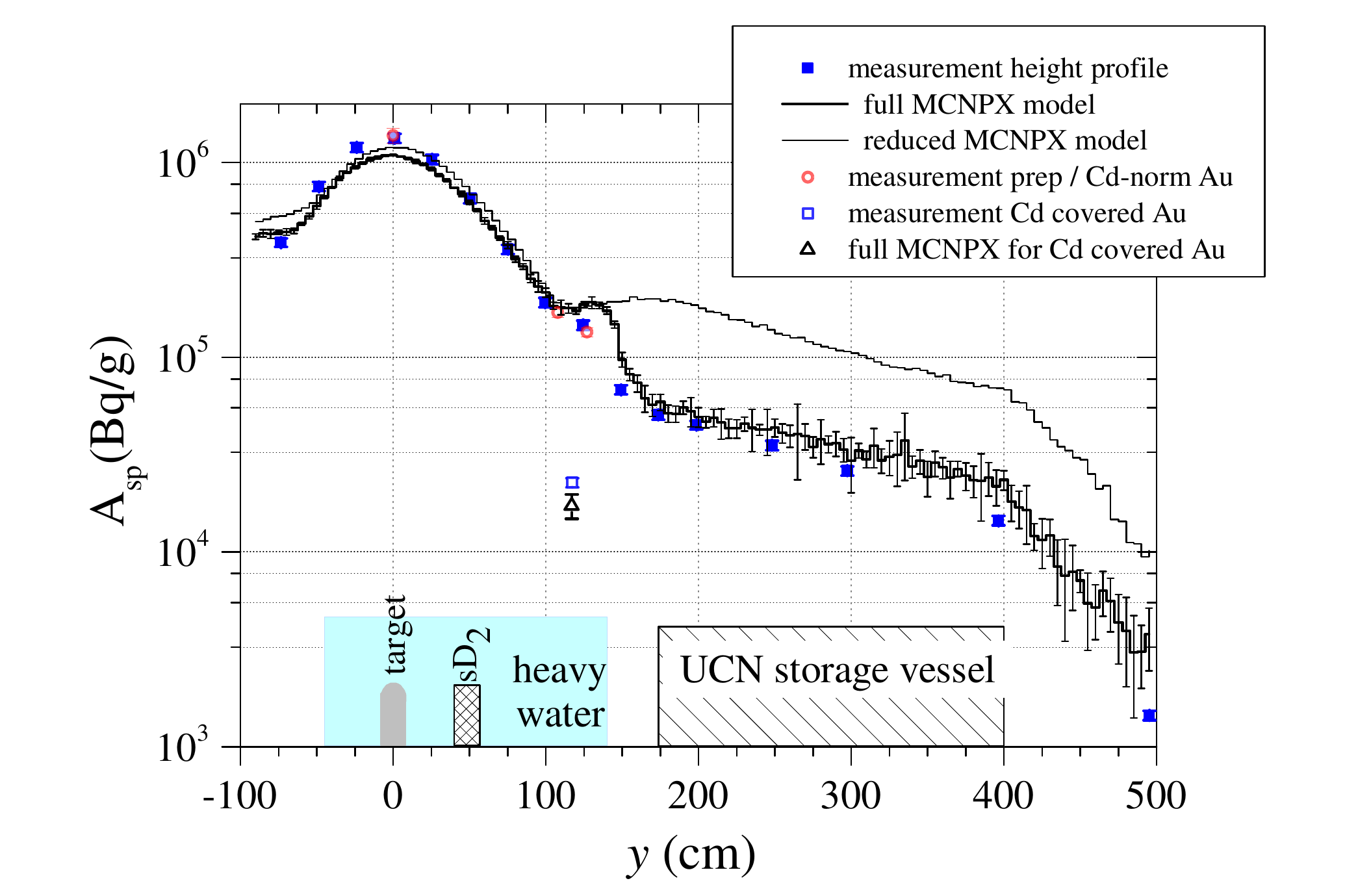}
\caption{
$^{198}$Au specific activity A$_{sp}$(\,Bq/\,g) versus height (y) along the UCN tank
after irradiation with
a 1\,s long beam pulse at 2.2\,mA proton beam current.
Comparison of simulations with the 
full MCNPX model (bold solid line) with statistical errors, and 
the reduced MCNPX model (thin solid line).
Measurement results from Tab.\ref{tab:result-h} are depicted as filled symbols
those from Tab.\ref{tab:result-cd} as open circles.
The measured activity for the Cd-covered Au foil is shown as open square,
the corresponding simulation as open triangle.
}
\label{results-comparison}
\end{figure*}

In Fig.~\ref{results-comparison} 
the results of the activity calculation for the simulated neutron flux
using the full and the reduced MCNPX
geometry model are compared to
the activity measurements provided in Tab.~\ref{tab:result-cd}
and \ref{tab:result-h}. 
The statistical error of the activity calculations 
between $-$100 and $+$150\,cm 
is below $\pm$4\%
for both geometry models.
Above 150\,cm the statistical error is $\sim$20\%
for the full model and $\sim$10\% for the calculation with the reduced MCNPX model.
This precision is completely sufficient for our purposes.

\begin{table}[h]
\centering
\begin{tabular}{|c|c|c|c|}\hline 
 & $a_\text{tot}$           &   $a_\text{epi}$   &  $a_\text{epi}$/$a_\text{tot}$ \\ 
 &   ($\,kBq/\,g$)          &   ($\,kBq/\,g$)    &                                \\  \hline
measurement  & 152$\pm$10   &   22.8$\pm$1.2     & 0.150$\pm$0.010                \\ \hline
calculation  & 175$\pm$10   &   17.3$\pm$2.4     & 0.100$\pm$0.015                 \\ \hline 
\end{tabular}
\caption{
Measured (from Tab.\ref{tab:result-cd}) 
and simulated specific $^{198}$Au activity from the total, $a_{tot}$, 
and epithermal, $a_\text{epi}$, neutron flux, 
and their ratio
(for the position of the Cd covered gold foil) 
at a height of 117\,cm.
The $a_\text{tot}$(measurement) value is interpolated 
between the two values of Tab.\ref{tab:result-cd}.
}
\label{Tab_40-01}
\end{table}

An additional MCNPX calculation was done in the full model 
in order to compare to the measurement with the cadmium (Cd) covered foil
at y=117\,cm
which is only sensitive to the epithermal neutron flux.
The results for the Cd covered gold foil 
are compared with the corresponding 
adjacent uncovered gold foils
in Tab.~\ref{Tab_40-01}
and are also shown in Fig.~\ref{results-comparison}.
The epithermal neutron contribution to the unshielded 
gold activity at the measurement 
position is only around 15$\%$. 
Therefore, even a large error on the 
knowledge of the epithermal neutron flux contributes 
relatively little to the error of the activity from the total neutron flux.

The crucial quantity of interest for the production of ultracold neutrons
is the thermal neutron flux entering the solid deuterium moderator vessel
located about 50\,cm above the proton beam plane.
As the thermal neutron flux at this specific position cannot be directly 
probed, one has to compare experiment and simulation results at accessible positions.
The measured $^{198}$Au activity at this height is 654$\pm$35\,kBq/g. It has to be compared with 
the values calculated with the full, 626$\pm$16\,kBq/g, and the reduced, 688$\pm$8\,kBq/g, 
MCNPX model. The two models differ only by about 10$\%$ at this position.
The very good agreement between measured and calculated activities
demonstrates that the combination of neutron production in the Cannelloni target 
and subsequent moderation in the surrounding D$_2$O are
accurately modeled by MCNPX and understood.
%


\section{Summary and discussion}
\label{summary}


%
The measurements of the gold activity along the 
UCN tank from 1\,m below 
to $\sim$3\,m above the proton beam plane are well matched
by the full MCNPX model 
(see Tab.~\ref{tab:simulation-result}).
A correct simulation of the thermal neutron flux 
along the height is only possible with a realistic 
description of structural details inside and
the presence of shielding outside of the tank.
For our ``full'' model the result of the simulation differs 
to measurements
up to 50$\%$ if one looks at heights of 4\,m above beam plane.
A more detailed modeling would be required for the top region 
to reach better matching but was discarded, as this is not required to 
understand the UCN production.

An assessment of the epithermal neutron flux at 117\,cm above the beam plane was
done with a comparison measurement of a cadmium covered gold foil.
The measurement shows a 15$\%$ fraction of epithermal neutrons at that position, while
the simulation gives a $\sim$ 10 \% fraction.
This difference contributes relatively little to the total activity value
and the determination of the thermal flux.

The primary region of interest 
for the production of ultracold neutrons 
is the location of the solid deuterium moderator vessel. 
The $^{198}$Au activity measurement at this beam height of about 50\,cm
falls between the values calculated with the full and the reduced 
MCNPX model, the two models differ by about 10$\%$ at this position.

The good agreement between measurements and simulations shows that 
the spallation target and D$_2$O moderator perform as expected
and that the 
basic processes of neutron production and moderation are 
understood in the specific geometry of the PSI UCN source.


\section{Acknowledgements}
We acknowledge the support of 
Max Ruethi, Michael Meier,
the BSQ group, 
especially Kurt Geissmann and August Kalt, 
Markus M\"ahr and the AMI workshops at PSI,
Othmar Morath and Pascal Meier of the radiation protection division,
Anton Mezger and the PSI accelerator operating crew,
and discussions with Sven Forss.
Support by the Swiss National Science Foundation
Projects 200020\_137664 and 200020\_149813 is gratefully acknowledged.

\hbadness=10000  


\end{document}